\documentclass[prl,aps,twocolumn,preprintnumbers,showpacs]{revtex4}
\usepackage{epsfig,amssymb,amsfonts}
\def\id{\mathbb{I}}
\newcommand{\ket}[1]{\left|#1 \right\rangle}
\newcommand{\bra}[1]{\left\langle #1\right|}
\newcommand{\proj}[1]{\ket{#1}\bra{#1}}

\def\opone{\leavevmode\hbox{\small1\kern-3.8pt\normalsize1}}
\newcommand{\tr}[1]{\mbox{Tr}#1 }
\newcommand{\vis}{\text{v}}
\newcommand{\hilb}{\mathcal{H}}

\begin{document}
\title{Interference of Quantum Channels}
\date{\today}
    \author{Daniel K. L. \surname{Oi}}%
    \affiliation{Centre for Quantum Computation,
    Department of Applied Mathematics and Theoretical Physics,
    University of Cambridge,
    Wilberforce Road,
    Cambridge CB3 0WA,
    United Kingdom}%

\begin{abstract}
  We show how interferometry can be used to characterise certain aspects of
  general quantum processes, and in particular, the coherence of completely
  positive maps. We derive a measure of coherent fidelity, the maximum
  interference visibility, and the closest unitary operator to a given physical
  process under this measure.
\end{abstract}


\maketitle

\section{Introduction}

A key requirement of quantum information processing is the ability to transform
states coherently~\cite{NC2000}. In general, quantum processes will be
described by quantum channels, or Completely Positive (CP)
maps~\cite{Kraus1983}. A relevant question is what happens when different
processes act simultaneously on a system. Surprisingly, knowledge of the
individual quantum channels alone is insufficient to specify the action of the
simultaneous operation of both maps~\cite{Aaberg2003}.  An experimental
determination of the interference of the two maps reveals additional
information about the maps which is not taken into account in their individual
descriptions and is a measure of their coherent properties.  From this, we may
define an operational definition of coherent fidelity between CP maps. Thus,
interferometry can be used a tool to extract information inaccessible to
conventional process tomography~\cite{CCL2001}.

\section{Interferometry}

\begin{figure}
\includegraphics[width=0.45\textwidth]{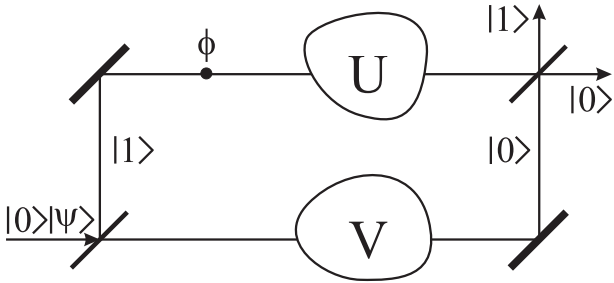}
\caption{Mach-Zender Interferometer.
  We allow the possibility of different quantum processes occurring in the
  upper and lower arms.}
\label{fig:machzender}
\end{figure}

Single particle interference (Fig.~\ref{fig:machzender}) displays the key
elements of quantum mechanics: the superposition of indistinguishable paths,
and the complementarity of certain observables. Interference is a consequence
of the possibility of the particle taking both paths, and any process which
tends to label the path of the particle will reduce the magnitude of the
interference~\cite{SKE1999}.

\begin{figure}
\includegraphics[width=0.45\textwidth]{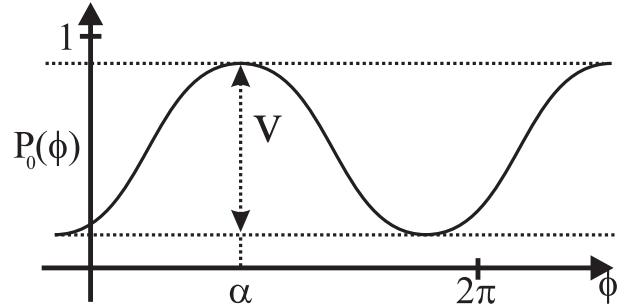}
\caption{Interference pattern showing a phase shift and reduction in visibility.
  The shift is a measure of the relative phase of the two quantum processes,
  and the reduction of visibility is a consequence of the leakage of path
  information into other degrees of freedom.}
\label{fig:pattern}
\end{figure}

\begin{figure}
  \includegraphics[width=0.45\textwidth]{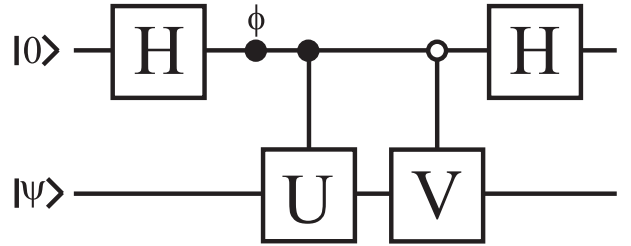}
\caption{Quantum Network for Interfering Unitaries. The actions of the unitary
  operations on the internal state of the particle (lower line) are controlled
  by the path of the particle (upper line).}
\label{fig:machnet}
\end{figure}

In general, the perfect interference pattern will be modified by the presence
of quantum processes occurring in the upper and lower arms
(Fig.~\ref{fig:pattern}). For unitary processes (Fig.~\ref{fig:machnet}), the
evolution of a particle with initial internal state $\ket{\psi}\in\hilb_d$ is,
\begin{eqnarray}
\ket{\Psi_{in}}&=&\ket{0}\ket{\psi}\\
&\mapsto&\frac{1}{\sqrt{2}}(\ket{0}+\ket{1})\ket{\psi}\nonumber\\
&\mapsto&\frac{1}{\sqrt{2}}(\ket{0}V\ket{\psi}+e^{i\phi}\ket{1}U\ket{\psi})\nonumber\\
&\mapsto&\frac{1}{2}\left[\ket{0}(V+e^{i\phi}U)\ket{\psi}+\ket{1}(V-e^{i\phi}U)\ket{\psi}\right].
\nonumber
\label{psi_out}
\end{eqnarray}
The probability of finding the particle in the $\ket{0}$ state, corresponding
to it exiting the interferometer from the horizontal output port, is given
by
\begin{eqnarray}
P_{0}(\phi)&=&\frac{1}{4}\bra{\psi}\left(V+e^{i\phi}U\right)^\dagger
\left(V+e^{i\phi}U\right)\ket{\psi} \nonumber \\
&=& \frac{1}{2}\left(1+\vis\cos(\phi-\alpha)\right) \label{P0}
\end{eqnarray}
where $|\vis|\equiv|\bra{\psi}U^{\dag}V\ket{\psi}|$ is the new visibility of
the interference pattern, and $\alpha \equiv
\arg\left(\bra{\psi}U^{\dag}V\ket{\psi}\right)$ is the shift of the
interference fringes~\cite{pancha56}. The magnitude of the visibility is the
fidelity of the states $U\ket{\psi}$ and $V\ket{\psi}$, i.e. the overlap of the
states exiting the upper and lower arms of the interferometer~\cite{EMOHK2002}.
The higher their fidelity -- hence the lower their distinguishability -- the
greater the interference effect.  Conversely, if the states exiting the upper
and lower arms were perfectly distinguishable (orthogonal), there would be no
interference.

If the initial state of the particle is
$\varrho_{in}=\ket{0}\bra{0}\otimes\rho$, the modified visibility is
\begin{eqnarray}
 {\vis}e^{i\alpha}=\tr\left[\rho U^{\dag}V\right], \label{vis}
\end{eqnarray}
which is the expectation value $\langle U^\dagger V \rangle_\rho$. If we use
the input $\rho=\frac{\id}{d}$, the maximally mixed state (equivalent to
randomly sampling over a uniform distribution of pure input states), then the
visibility pattern gives us the quantity
\begin{eqnarray}
\tr\left[U^{\dag}V\right]=d {\vis}e^{i\alpha}, \label{vis_mixed}
\end{eqnarray}
from which the Hilbert-Schmidt distance between two operators on $\hilb_d$ can
be derived,
 \begin{eqnarray}
 D^2(U,V) &=& \tr((U-V)^{\dag}(U-V))\nonumber\\
 &=&2\left(d - Re\left\{\tr\left[U^{\dag}V\right]\right\}\right)\nonumber\\
&=&2d(1-\vis\cos\alpha). \label{dist}
\end{eqnarray}
Hence, we have a direct estimate of the distance between unitary processes~\cite{OARF2003}.

\section{CP Maps}

\begin{figure}
\includegraphics[width=0.45\textwidth]{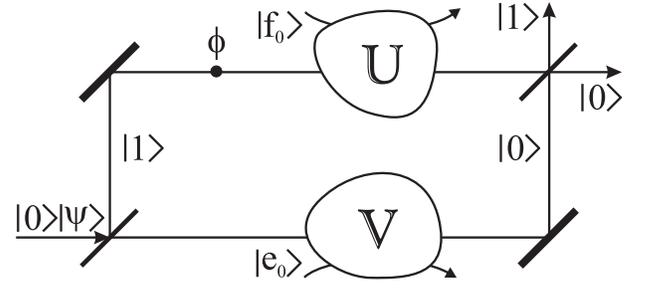}
\caption{Interference of two CP maps. We model the situation by implementing
  $\mathcal{U}$ and $\mathcal{V}$ by unitaries $\mathbb{U}$ and $\mathbb{V}$
  respectively acting on larger Hilbert spaces, and then tracing out the
  ancillary systems.}
\label{fig:cpmap}
\end{figure}

\begin{figure}
\includegraphics[width=0.45\textwidth]{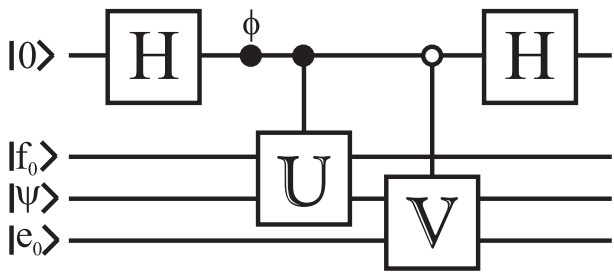}
\caption{Quantum network for the interference of two CP maps implemented
  by unitary operators acting upon the system and ancillas.}
\label{fig:cpnet}
\end{figure}

The interference pattern can reveal important information in the case where the
operations in the upper and lower arms are not unitary, but are CP maps
$\mathcal{U}$ and $\mathcal{V}$. We will assume these are trace preserving and
have the same input and output (finite dimensional) spaces. We can model this
case by extending the state space of the entire system by appending two
ancillas F and E (assuming that $\mathcal{U}$ and $\mathcal{V}$ are independent
processes), which are coupled to the top and bottom paths by overall unitaries,
$\mathbb{U}$ and $\mathbb{V}$, which implement the CP maps, $\mathcal{U}$ and
$\mathcal{V}$ respectively~\cite{Stinespring1955} (Figs.~\ref{fig:cpmap} \&
\ref{fig:cpnet}),
\begin{eqnarray}
\mathcal{U}(\rho)&=&\tr_{F}\left[\mathbb{U}(\rho\otimes\proj{f_0})\mathbb{U}^\dagger\right]\\
\mathcal{V}(\rho)&=&\tr_{E}\left[\mathbb{V}(\rho\otimes\proj{e_0})\mathbb{V}^\dagger\right],
\end{eqnarray}
where $\{\ket{e_\mu}\}$ and $\{\ket{f_\nu}\}$ are orthonormal bases and
$\ket{e_0}$ and $\ket{f_0}$ are initial states of E and F
respectively.

Note that for any CP map $\Lambda$, there exists many unitaries which
implement $\Lambda$ and that we cannot distinguish by quantum process
tomography the different instantiations. We may instead uniquely specify
$\Lambda$ via the Jamio{\l}kowski isomorphism~\cite{Jamiolkowski1972},
\begin{equation}
\Lambda\cong \varrho_\Lambda\equiv\id\otimes\Lambda(\proj{\Psi_+}),
\label{eq:jamiolkowski}
\end{equation}
where $\ket{\Psi_+}$ is the maximally entangled state and there exists a
one-one correspondence between the set of CP maps and the set of bipartite
density operators having one subsystem with maximally mixed reduced density
operator.

We can trivially extend both the overall unitaries to act on the whole space
of the particle, and ancillas E and F,
\begin{eqnarray}
\tilde\mathbb{U}&=&\mathbb{U}\otimes\id_E\\
\tilde\mathbb{V}&=&\mathbb{V}\otimes\id_F.
\end{eqnarray}
The action of the interferometer on an initially pure state of the particle is
now given by
\begin{eqnarray}
\ket{\Psi}&=&\ket{0}\ket{\psi}\ket{e_0}\ket{f_0}\\
&\mapsto&\frac{\ket{0}\tilde\mathbb{V}\ket{\psi e_0 f_0}
+e^{i\phi}\ket{1}\tilde\mathbb{U}\ket{\psi e_0 f_0}}{\sqrt{2}}\nonumber\\
&\mapsto&\frac{\left(\ket{0}(\tilde\mathbb{V}+e^{i\phi}\tilde\mathbb{U})
+\ket{1}(\tilde\mathbb{V}-e^{i\phi}\tilde\mathbb{U})\right)\ket{\psi e_0 f_0}}{2}.\nonumber
\end{eqnarray}

Thus the probability of the particle exiting from the horizontal output port
is given by
\begin{equation}
\text{P}_{0}(\phi)=
\frac{1}{4}\left|\left(\tilde\mathbb{V}+e^{i\phi}\tilde\mathbb{U}\right)
\ket{\psi e_0 f_0}\right|^2.
\end{equation}
In general, if the particle has internal state $\rho$, the probability is
\begin{equation}
\text{P}_0(\phi)=\frac{1}{2}\left(1+Re\left\{e^{i\phi}
\tr\left[\tilde\mathbb{U}^\dagger\tilde\mathbb{V}\rho\otimes\proj{e_0 f_0}\right]\right\}\right).
\end{equation}
The relevant quantity in the above can be expressed as
\begin{equation}
\tr\left[\tilde\mathbb{U}^\dagger\tilde\mathbb{V}\varrho\otimes\proj{e_0 f_0}\right]
=\tr\left[\upsilon_0^\dagger \nu_0 \varrho\right],
\end{equation}
where the Kraus operators for $\mathcal{U}$ and $\mathcal{V}$ are given by
\begin{eqnarray}
\left\{\upsilon_i\right\}&=&\left\{\bra{e_i}\mathbb{U}\ket{e_0}\right\}\\
\left\{\nu_j\right\}&=&\left\{\bra{f_j}\mathbb{V}\ket{f_0}\right\}.
\end{eqnarray}
Again, if the input is the maximally mixed state, the interference pattern
depends on $\frac{1}{d}\tr[\upsilon_0^\dagger \nu_0]$ only. This reduces to the
previously considered case if the operations in the upper and lower paths are
unitary on the internal state of the particle.

It is interesting to note that the visibility is dependent on a particular
decomposition of the two CP maps, in particular the overlap of the first Kraus
operators of $\mathcal{U}$ and $\mathcal{V}$. We may interpret this in the
framework of quantum jumps~\cite{ESBOP02,faria02}. The visibility is a
consequence of the indistinguishability of the two possible paths of the
particle through the interferometer, anything that serves to tag the passage
of the particle serves to reduce the interference pattern. This may not just
be an internal change in the state of the particle (created by differing
unitary operations $U$ and $V$) but also any changes in the environment which
may mark the particle's passage. Thus, the first Kraus operators of both
$\mathcal{U}$ and $\mathcal{V}$ denote the action of the operation when there
is no quantum jump of the environment ``watching'' the particle. The residual
overlap concerns the ``unitary'' action of the CP map under this condition of
no jump.

Note that even though a CP map may seem to be unitary when only acting on the
internal degrees of freedom, i.e.
\begin{equation}
\Lambda(\rho)=U\rho U^\dagger,
\end{equation}
the first Kraus operator may be the zero operator,
$\{\lambda_0=0,\lambda_1=U\}$, and hence give zero visibility~\footnote{Thanks
  to B-G Englert for pointing out this subtle case.}, e.g.
\begin{equation}
\mathbb{U}(\ket{\psi}_{I}\ket{f_0})=(U\ket{\psi}_{I})\ket{f_1}\, \forall \ket{\psi}\in\hilb_d.
\label{eq:notunitary}
\end{equation}
In this case, the map serves as an indicator of path, entangling the fact of
the passage of the particle with the environment without altering the internal
state.  In the interferometer, this results in the destruction of all
interference. In general, a CP map will necessarily entangle the passage of the
particle with environmental degrees of freedom (a non-product $\mathbb{U}$),
hence reducing the interference beyond the effect of altering the internal
state of the particle.

We may thus define a coherent fidelity between two sets of Kraus operators
implementing different CP maps as,
\begin{equation}
\mathcal{F}\left(\{\upsilon_i\},\{\nu_i\}\right)
=\frac{1}{d}\left|\tr\left[\upsilon_0^\dagger\nu_0\right]\right|,
\end{equation}
and their relative phase,
\begin{equation}
\mathcal{P}\left(\{\upsilon_i\},\{\nu_i\}\right)
=\arg\left(\tr\left[\upsilon_0^\dagger\nu_0\right]\right ).
\end{equation}
We may note that another fidelity measure on the set of CP maps has
been defined via the Uhlmann fidelity~\cite{Uhlmann1976} between the density
operators defined in Eq.~(\ref{eq:jamiolkowski})~\cite{Raginsky2001},
\begin{equation}
\mathcal{F}^{'}\left(\mathcal{U},\mathcal{V}\right)
=\tr\left[\sqrt{\sqrt{\rho_{\mathcal{U}}}\rho_{\mathcal{V}}\sqrt{\rho_{\mathcal{U}}}}\right].
\end{equation}

\section{Maximally Coherent CP Maps}

For a set of Kraus operators $\{\lambda_i\}$ defining a CP map $\Lambda$, we
may define a measure of its self-coherence by inserting two independent
instances of $\Lambda$ into both arms of an interferometer. If the CP map is
unitary, then the interference pattern will have unit visibility. However, if
there are more than one Kraus operator,
\begin{equation}
\vis=\frac{1}{d}\tr\left[\lambda_0^\dagger\lambda_0\right]<1,
\end{equation}
we can take this to measure the distance of $\Lambda$ to the set of unitaries.

It is interesting to ask, for a given $\Lambda$, and for all possible
compatible sets of Kraus operators, what is the maximum visibility or
self-coherence? In other words, for all sets of Kraus operators $\{\lambda_i\}$
implementing $\Lambda$, what is the largest value of
$\frac{1}{d}\tr[\lambda_0^\dagger\lambda_0]$?  The canonical method of
constructing a set of Kraus operators of a CP map is given by
Choi~\cite{Choi1975}. The operators $\{\lambda_i\}$ created are linearly
independent and thus represent the minimum number of operators required to
represent $\Lambda$. If $\{\lambda_i^{'}\}$ also implement $\Lambda$, they are
related by
\begin{equation}
\lambda_i^{'}=\sum_k u_{ik}\lambda_k,\label{eq:isom}
\end{equation}
where $u$ is an isometry in general, or unitary when the number of elements in
each set are equal. It can easily be shown that the largest possible
visibility is obtained when $\{\lambda_i\}$ are orthogonal,
\begin{equation}
\tr\left[\lambda_i^\dagger \lambda_j\right]
=\delta_{ij}\tr\left[\lambda_j^\dagger\lambda_j\right],
\label{eq:orthog}
\end{equation}
and the largest element is $\lambda_0$. When $\Lambda$ is placed in both arms,
this upper limit is a measure of the intrinsic (de)coherence of the process.
The visibility of an actual realisation of $\Lambda$ may be smaller than this
maximum due to processes as in Eq.~(\ref{eq:notunitary}) but this does not
represent intrinsic decoherence of the map itself.

We can also find the closest unitary operator to a given set of Kraus
operators $\{\lambda_i\}$ which induce the CP map $\Lambda$ by considering an
interferometer with $\Lambda$ occurring one arm, and a unitary $U$ operation in
the other arm which we may alter as we like. We can maximise the interference
pattern by changing $U$, and hence obtain the closest unitary to
$\{\lambda_i\}$. The visibility is given by
\begin{equation}
\vis=\frac{1}{d}\left|\tr\left[\lambda_0^\dagger U\right]\right|
=\frac{1}{d}\left|\tr\left[\sqrt{\lambda_0 \lambda_0^\dagger}U_{\lambda_0} U\right]\right|,
\end{equation}
where we have used the polar decomposition of $\lambda_0=\sqrt{\lambda_0
  \lambda_0^\dagger}U_{\lambda_0}$~\cite{HJ1985}. The visibility is thus
maximised when $U=U_{\lambda_0}^\dagger$. If the eigenvalues of
$\lambda_0^\dagger \lambda_0$ are $\{r_j\}$, then the visibility when
$\Lambda$ is in both arms is simply $\vis_{\Lambda\Lambda}=\sum r_j$, whereas
if we compare $\Lambda$ with its closest unitary, it is $\vis_{\Lambda
  U}=\sum \sqrt{r_j}$, and it is easy to see that $\vis_{\Lambda
  U}\ge\vis_{\Lambda\Lambda}$.

We can also consider what is the maximum coherent fidelity between two CP maps
$\mathcal{U}$ and $\mathcal{V}$. Let $\{\upsilon_i\}$ and $\{\nu_j\}$ be
orthogonal Kraus sets for $\mathcal{U}$ and $\mathcal{V}$ respectively
(Eq.~(\ref{eq:orthog})), and $(A_{ij})=(\tr[\upsilon_i^\dagger \nu_j])$ be the
matrix of their inner products. If $\{\upsilon_i^{'}=\sum_j g_{ij}\upsilon_j\}$
and $\{\nu_i^{'}=\sum_j h_{ij}\nu_j\}$ are also compatible Kraus operators for
$\mathcal{U}$ and $\mathcal{V}$ (Eq.~(\ref{eq:isom})), then
\begin{equation}
\left|\tr\left[\upsilon_0^{'\dagger}\nu_0^{'}\right]\right|
=\left|
\sum_{ij} \tr\left[g_{0i}^{*}\upsilon_i^\dagger h_{0j}\nu_j\right]
\right|
=\left|
\sum_{i} g_{0i}^{*} \sum_{j} A_{ij} h_{0j}
\right|
\end{equation}
is maximised when
\begin{eqnarray}
\left\|A\vec{g}_{0}\right\|&=&\max_{\|\vec{g}\|=1} \left\|A\vec{g}\right\|\\
\vec{h}_{0}&=&\frac{A\vec{g}_{0}}{\left\|A\vec{g}_{0}\right\|},
\end{eqnarray}
where we have used the operator norm of $A=(A_{ij})$, and $\vec{g}_{0}$ and
$\vec{h}_{0}$ are the first column vectors of the isometries $g_{ij}$ and
$h_{ij}$ relating $\{\upsilon_i\}$ and $\{\nu_j\}$ to $\{\upsilon_i^{'}\}$ and
$\{\nu_i^{'}\}$ respectively. This reduces to the previous case where both CP
maps are the same and thus $(A_{ij})$ is real diagonal.

\section{Conclusion}

Interferometry can be applied to the case of non-unitary processes to extract
information about the underlying physical processes which implement them. In
particular, we can derive a measure of the coherence of a quantum operation,
its maximum for any CP map, and the closest unitary under this measure. It is
to seen whether consideration of dynamical CP maps can impose further internal
structure on quantum operations.

\begin{acknowledgments}
  I would like to acknowledge the support of the Cambridge-MIT Institute
  Quantum Information Initiative, EU grants RESQ (IST-2001-37559) and TOPQIP
  (IST-2001-39215). I also thank S. G. Schirmer, A. K. Ekert, A. P. A.  Kent,
  A.  Landahl, B-G.  Englert, D. Kaszlikowski, L. C. Kwek, A. J. Short and B.
  M. J.  B. D. Walker for helpful discussions.
\end{acknowledgments}


\begin{thebibliography}{99}
\bibitem{NC2000}M. A. Nielsen, I. L. Chuang, {\em Quantum Computation and
    Quantum Information}, Cambridge University Press (2000)
\bibitem{Kraus1983}K. Kraus, {\em States, Effects, and Operations},
  Springer-Verlag, Berlin (1983)
\bibitem{Aaberg2003}J. {\AA}berg, electronic pre-print quant-ph/0302182
\bibitem{CCL2001} A. M. Childs, I. L. Chuang, D. W. Leung, Realization of
  Quantum Process Tomography in NMR, {\em Phys. Rev. A.} {\bf 64}, art. no.
  012314 (2001)
\bibitem{SKE1999} P. D. D. Schwindt, P. K. Kwiat, B-G. Englert, Quantitive
  wave-particle duality and non-erasing quantum erasure, {\em Phys. Rev. A}
  {\bf 60}, p4285 (1999)
\bibitem{pancha56}S. Pancharatnam, {\em Proc. Indian Acad. Sc.}
  {\bf 44}, p247 (1956)
\bibitem{EMOHK2002} A. K. Ekert, C. Moura Alves, D. K. L Oi, M.
  Horodecki, P. Horodecki, L. C. Kwek, Direct Estimations of Linear and
  Nonlinear Functionals of a Quantum State, {\em Phys. Rev. Lett.} {\bf 88},
  art. no. 217901 (2002)
\bibitem{OARF2003}D. K. L. Oi, D. Angelakis, R. Rodriquez, B. Falabretti,
  Forthcoming
\bibitem{Stinespring1955}W. F. Stinespring, Positive Maps on C$^*$-Algebras,
  {\em Proc. Amer. Math. Soc.} {\bf 6}, p211 (1955)
\bibitem{ESBOP02}M. Ericsson, E. Sj\"oqvist, J. Br\"annlund, D. K. L. Oi, A.
  K. Pati, quant-ph/0205160
\bibitem{faria02}J. G. Peixoto de Faria, A. F. R. de Toledo Piza, M. C. Nemes,
  quant-ph/0205146
\bibitem{Jamiolkowski1972} A. Jamio{\l}kowski, \emph{Rep. Math. Phys.} {\bf 3}, p275 (1972)
\bibitem{Uhlmann1976}A. Uhlmann, The transition probability in the state space
  of a *-algebra, {\em Rep. Math. Phys.} {\bf 9}, p273 (1976)
\bibitem{Raginsky2001}M. Raginsky, A Fidelity Measure for Quantum Channels,
  {\em Phys. Lett. A} {\bf 290}, p11 (2001)
\bibitem{Choi1975}M. D. Choi, {\em Linear Algebra and its Applications} {\bf
    10}, p285 (1975)
\bibitem{HJ1985}R. A. Horn, C. R. Johnson, {\em Matrix Analysis}, Cambridge
  University Press (1985)
\end{thebibliography}
\end{document}